\documentclass[10pt, reprint, aps, amsmath,amssymb,twocolumn,article]{revtex4-1}

\usepackage{epigraph}
\usepackage{graphicx}
\usepackage{dcolumn}
\usepackage{algorithm,algpseudocode}
\usepackage{bm}
\usepackage[font=scriptsize]{caption}
\usepackage{booktabs}
\usepackage{subcaption,graphicx}
\algnewcommand{\To}{\textbf{To }}
\algnewcommand\Input{\item[\textbf{Input:}]}%
\algnewcommand\Output{\item[\textbf{Output:}]}%

\begin{document}

\title{Analysis of the Common Emitter Amplifier Taking into Account Transistor Non-Linearity}

\author{Luciano da F. Costa}
\email{luciano@ifsc.usp.br, copyright LdaFcosta}
\affiliation{S\~ao Carlos Institute of Physics, IFSC-USP,  S\~ao~Carlos, SP,~Brazil}

\date{\today}

\begin{abstract}
The operation of a typical common emitter amplifier, including negative feedback, is studied taking into account the 
non-linearity characteristic of real-world transistors.  This has been accomplished by employing a recently proposed Early 
modeling approach, which allowed the analytical equations to be obtained describing the current and voltage behavior in the 
adopted common emitter circuit.  Average and dispersion (coefficient of variation) of the current and voltage gains can
then be calculated and used to characterize the common emitter amplification while reflecting the transistor non-linearity.
Several interesting results were obtained, including the fact that the negative feedback provided by the emitter
resistance is not capable of completely eliminating effects of parameters differences exhibited by transistors.
Importantly, transistors with larger Early voltage $V_a$ magnitudes tended to provide significantly enhanced linearity even when
substantial negative feedback is used. These results motivates customized design, implementation and application approaches
taking into account the parameters of the available devices.
\end{abstract}

\keywords{Amplifiers, common emitter, Early effect, Early model, distortion, theoretical circuit analysis, 
circuit design, equivalent circuits, gain, linearity.}
\maketitle

\setlength{\epigraphwidth}{.49\textwidth}
\epigraph{``$\mu \eta \delta \grave{\epsilon} \nu ~  \ddot{\alpha} \gamma \alpha \nu.$'}{Delphic Proverb}

\section{Introduction}

The common emitter design is one of the most fundamental amplifying circuit configurations, representing a reference
in electronics circuits be it for its central role in research, application, or education.  This circuit configuration can be found in
virtually almost every textbook on electronics circuits 
(e.g.~\cite{pettit:1961,gray:1969,gronner:1970,sedra:1998,jaeger:1997,boylestad:2008,horowitz:2015,streetman:2016}). 
Consequently, much has been said and written about this 
configuration, including its electronic properties as voltage gain, input and output resistance, linearity/distortion, the negative 
feedback action implemented by the emitter resistance, stability with component  variations, frequency response, to name a 
few among a set of issues encompassing a great deal of the main interests in electronics.  
This circuit typically includes a voltage source attached through a capacitor to voltage divider driving the amplifier input port, 
an emitter resistance for  negative feedback shared by both the input and output mesh, and a resistance load in series with 
an external voltage supply.  Either the input voltage or current are taken as the input signal, while the collector voltage, or 
that across the load, are often understood as the output signal.  

The main purpose of the common emitter amplifier is to produce current and/or voltage gain over the load or output, while
affecting as little as possible the signal shape.
The intrinsic parameter variability exhibited by real-world transistors, allied to the need to improve amplification linearity and
stability, motive the incorporation of the emitter resistance as a means to implement negative feedback.  However, it is also 
possible (at the expense of some stability loss) to consider variations of the classic common emitter configuration devoid of the 
emitter resistance, which can be  considered as a means to maximize gain and achieve wider voltage excursion across the 
load (i.e.~avoid the voltage drop  at the emitter resistance).  

Typically, studies of the common emitter circuit involve the adoption of a model and respectively implied equivalent circuit
to represent the involved transistor.  The most often employed approaches rely on two electronically meaningful parameters, 
namely the current gain $\beta$ and the output resistance $R_o$.  This type of approach, however, cannot take into account the
nonlinearities that are unavoidably found in any real-world transistor.  These nonlinearities are a direct consequence of the 
fanned structure of the base current ($I_B$) indexed characteristic isolines of the transistor behavior along its operation 
space $(V_C,I_C)$, where $V_C$ ad $I_C$ stand for the collector voltage and current.  

The above described geometrical organization of transistors action is a sole and direct consequence
of the \emph{Early effect} (e.g.~\cite{early:1952,jaeger:1997,streetman:2016}), in which the base width varies in terms of the 
voltage applied to  the base-collector junction.
This effect implies mots of the $I_B$-indexed characteristic isolines to cross at the same point $(V_C = V_a, I_C = 0)$ along the 
$V_C$ axis.  An immediate implication is that the collector output resistance will depend on $I_B$.  Another critically induced
property regards the fact that the local current gain $\beta$, as well as the output resistance $R_o$ become a function 
of \emph{both} $I_B$ and $V_C$.  Consequently,
transistor modeling approaches relying on $\beta$ and $R_o$ as parameters need to consider their respective averages or point 
values. Frequently, it is assumed that the fanned isolines can be approximated by a series of equispaced, parallel isolines with
the same inclination small $m = 1/R_o(I_B)$.  As such, this simplified device is typically represented by an equivalent circuit
involving a variable current source in parallel with a fixed output resistance. As an inevitable consequence, this transistor model 
becomes intrinsically linear, which is a severe simplification of what really happens in real-world transistors. 

Recently~\cite{costaearly:2017,costaearly:2018,germanium:2018,costaequiv:2018}, an alternative transistor model was proposed
that is simple and yet can accurately account for the representation of the transistor non-linearities induces by the fanned
(or radiating) characteristic isolines.  In addition, the two parameters used in this model, namely the Early voltage $V_a$ and a
proportionality constant $s$, are constant for any given device, not depending of $I_C$ or $V_C$.  These distinguishing
features are allowed by taking into account the very Early effect that causes the fanned isolines as the geometrical basis of the model.
Actually speaking, this model also relies on another hypothesis, verified experimentally for hundreds of small signal 
BJTs~\cite{costaearly:2017,germanium:2018,costaearly:2017}, that
the angles $\theta$ of the radiating isolines with the horizontal ($V_C$) axis are directly proportional to $I_B$ through the proportionality
constant $s$, i.e.~$\theta = s I_B$, implying that $R_o = 1/tan(s I_B)$.  As $I_B$ is usually very small, in the order to $\mu A$, 
it is often possible to make $tan(s I_B) = 1/(s I_B)$ with minimal deviation, so that $Ro(I_B) = 1/(s I_B)$.  

Despite its recent introduction, the Early model has already been successfully applied to achieve more complete and accurate 
modeling and studies of: stability with respect to voltage supply oscillations~\cite{costaequiv:2018}, parallel combinations of 
transistors~\cite{costaequiv:2018}, characterization of the output resistance and current gain of silicon and germanium NPN 
and PNP small signal transistor  models~\cite{costaearly:2018, germanium:2018}, study of a simplified common emitter 
configuration~\cite{costaearly:2018}, investigation of complementary transistor pairs in push-pull circuits~\cite{costaearly:2018}, 
as well as the characterization of the effects of reactive loads in amplification~\cite{costareact:2018}, among other results.  
The Early approach has also allowed the identification of the
fact that the amplification in a simplified common emitter amplifier tends to be perfectly linear for very small resistive 
loads~\cite{costareact:2018}. In addition, it has been found~\cite{costaearly:2018,costaearly:2018,costareact:2018} that the
linearity of the transistors amplification devoid of feedback tends to depend only of $s$ (directly), and not of $V_a$.  
As a matter of fact, the simplicity and
completeness of the Early model in being able to represent in an inherently suitable geometrical way the non-linearities of 
real-world transistors pave the way to a large number of implications for transistor amplification and even linear electronics
in general.

The current work focuses on substantially extending previous analyses~\cite{costaequiv:2018,costareact:2018} that had addressed
a simplified version of the common emitter configuration.   In those works, the emitter was connected directly to ground, and
there was no voltage divider for polarization at the input port, so no negative feedback can be achieved.  Though that 
configuration was still representative of several characteristics of transistor amplification, leading to the identification of a series 
of interesting effects as listed above, it did not account for the \emph{negative feedback} afforded by the inclusion of an 
emitter resistor $R_E$.  This is very important
because much of modern electronics has relied on negative feedback, introduced by H. S. Black~\cite{black:1934},
in order to achieve improved linearity and stability to factors such as temperature and transistor parameter variations.  
These advantages are achieved at the expense of substantial
gain expense.  Yet, a recent study~\cite{costafeed:2017,costaearly:2018} considering several types of commercial small signal BJTs 
suggested that negative feedback may not be enough, in several situations, to completely eliminate the relatively large 
parameter variations exhibited by real-world devices.  It was also shown that negative feedback can introduce unwanted
harmonics for certain types of non-linearities.  

The key role of negative feedback and the common emitter amplifier in electronics, allied to the above mentioned recent 
findings suggesting some respective limitations, motivate a reappraisal of the classic common emitter amplifier configuration.  
Of particular interest would be to investigate the efficacy of negative feedback in minimizing transistor non-linearity with 
respect to different transistor characteristics.  For instance, will feedback work more effectively for transistor with large or 
small current gains $\beta$ or by choosing different values for the external circuit components?  If such differences can be
found, how much can feedback action be improved by starting with a more linear transistor?  How will these effects vary
with the intensity of the negative feedback, as set-up by the external resistors? These correspond to key 
issues in both discrete and integrated circuit analysis, design, implementation and applications.   These questions provide 
the main motivation and objectives of the present work.

We start by presenting the ``classic'' common emitter configuration, and then obtain a respective equivalent circuit by using
the Early modeling methodology.  Accurate mathematical expressions are obtained that characterize the voltages, currents and 
gains while taking into account the non-linear behavior of real-world transistors.  By using these equations, it becomes possible
to investigate in an objective, more complete, and accurate way how the transistor inherent non-linearity influences the
efficiency of negative feedback in promoting linear amplification.  This is done by considering the voltage gain average and
coefficient of variation (closely related to the level of non-linearity in the amplification)  A number of interesting respective 
results are  obtained and discussed, and the work is concluded by presenting some possibilities for further related research.

\section{The Classic Common Emitter Configuration}

\begin{figure}[h!]
\centering{
\includegraphics[width=7cm]{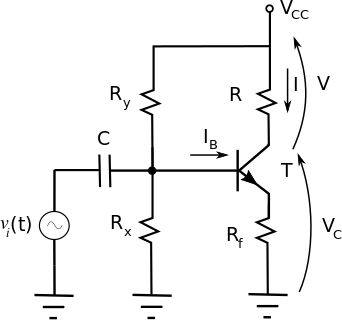}
\caption{The ``classic'' common emitter configuration as adopted in this work. The base-collector junction of transistor $T$ is
inversely biased, while the junction base-emitter is forward biased.  $R$ is the resistive load, and $R_f$ provides negative
feedback.  The input port is biased into an operation (quiescent) point $Q$ by the voltage divider implemented by $R_x$ and $R_y$,
receiving the input through the coupling capacitor $C$, which is chosen so as to have low impedance relatively to
the input bandwidth.}
\label{fig:classic}}
\end{figure}

Figure~\ref{fig:classic} shows the configuration of what we will henceforth refer to as the ``classic'' common emitter circuit.  A voltage
divider (resistors $R_a$ and $R_b$) is used to bias the input port, which is driven by an external signal voltage source $v_i(t)$,
attached through the decoupling capacitor $C$.  
The load is connected between the collector and the external voltage supply $V_{CC}$, while resistor $R_f$ is connected
between the emitter and ground in order to provide negative feedback.  In this circuit, the base-collector junction of transistor $T$ is
inversely biased, while the junction base-emitter is dforward biased.

First, we derive the equations for the input port, assuming that the forward-biased base-emitter junction is modeled
as a diode with offset voltage $V_r$ and resistance $R_r$.   By using Thevenin's theorem, we obtain the respective
equivalent circuit for the input voltage and divider as:

\begin{eqnarray}
   R_b = \frac{R_x R_y}{R_x + R_y}  \label{eq:Rb}  \\
   V_b = R_x\frac{V_{CC}}{R_x + R_y}  \label{eq:v}
\end{eqnarray}

\begin{figure}[h!]
\centering{
\includegraphics[width=9cm]{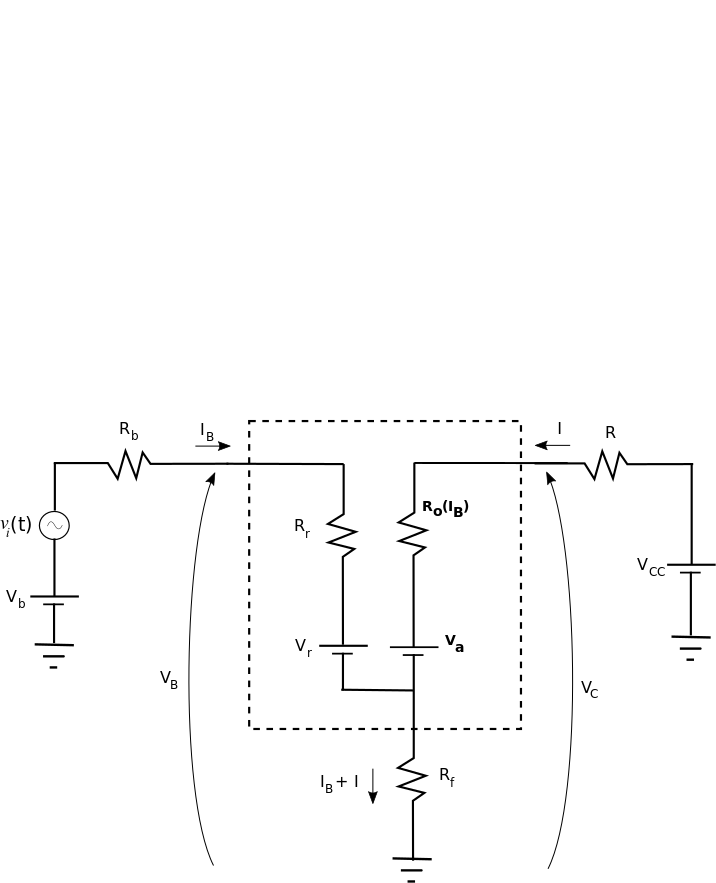}
\caption{The equivalent circuit of the common emitter configuration in Figure~\ref{fig:classic} obtained by
replacing the voltage divider in the input port by its Thevenin's equivalent circuit, and by replacing the output
mesh of transistor $T$ by the respective Early equivalent circuit.}
\label{fig:equiv}}
\end{figure}

Figure~\ref{fig:equiv} shows the common emitter circuit in Figure~\ref{fig:classic} modified to show the input port equivalent
circuit as well as the output mesh, with the transistor replaced by its Early equivalent circuit.
Observe that the input voltage source $v_i(t)$ is not considered as it acts differentially in the amplification, oscillating
around the operation point $Q$.  
As $R_b/(R_b+R_r)$ is typically very close to $1$ ($R_b$ is much larger than $R_r$), it becomes possible  to include 
$v_i(t)$ in series along the input mesh, for simplicity's and generality's sake, at the expense of nearly negligible error.

\begin{figure}[h!]
\centering{
\includegraphics[width=9cm]{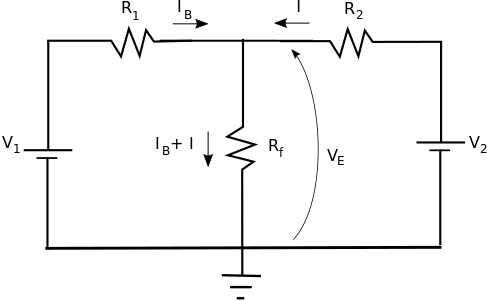}
\caption{The common emitter equivalent circuit in Figure~\ref{fig:equiv} can be further simplified into the depicted
configuration by making $R_1 = R_b + Rr$, $R_2 = R + R_o(I_B)$, $V_1 = v_i(t) + V_b - V_r$ and $V_2 = V_{CC} - V_a$.}
\label{fig:simplif}}
\end{figure}

For simplicity's sake, the above circuit can be further simplified to the configuration shown in Figure~\ref{fig:simplif}.
This has been achieved by subsuming the input and output series resistances and voltage sources, i.e.~$R_1 = R_b + Rr$,
$R_2 = R + R_o(I_B)$, $V_2 = V_{CC} - V_a$, and, in particular:

\begin{equation}
  V_1 = v_i(t) + V_b - V_r  \label{eq:V1}
\end{equation}

Despite its seeming symmetry and simplicity, this circuit exhibits some relative complexity as a consequence of $R_2$ being 
a function of $I_B$, i.e.~$R_2(I_B) = R + R_o(I_B) = R + 1/tan(s I_B) \approx R + 1/(s I_B)$. The variables 
in this circuit are $V$, $I_B$ and $I$.  By applying Kirchhoff's and Ohm's law, we have:

\begin{eqnarray}
\left\{ 
\begin{array}{c}
   V_1 - R_1 I_B - R_f (I + I_B) = 0 \\ 
   V_2 - R_2 I     - R_f (I +I_B) = 0   
\end{array}  \label{eq:system}
\right.
\end{eqnarray}

The solution for $I_B$ and $I$ in terms of $v_i(t)$ are given as $I_1$ and the DC current gain $A_i$, as in Equations~\ref{eq:I} 
and~\ref{eq:Ai}.

\begin{table*}
\begin{eqnarray}
   I_B(t) = \frac{s V_1 (R + R_f) - R_1 - R_f - s R_f V_2 - \aleph}{2 s (R R_1+R R_f+R_1 R_f)}   \label{eq:I} \\
   A_i(t) = \frac{I(t)}{I_B(t)} =     \frac{1}{2 R_f} \left( s R_f V_2 - s V_1 (R + R_f) -R_1 - R_f - \aleph \right)  \label{eq:Ai}  \\
   with \; \aleph = \sqrt{ 4 s V_1(R R_1 +  R R_f +  R_1 R_f)  +  (R_1 + R_f - s R V_ 1  + s R_f( V_2 - V_1)) ^2}
\end{eqnarray}
\end{table*}

These equations describe the operation, in terms of $V_1$ (and hence $v_i(t)$, through Equation~\ref{eq:V1}), of 
classic common emitter circuit as in Figure~\ref{fig:classic}.    

Proper operation of this circuit requires 
the conditions in Equation~\ref{eq:geq0} and~\ref{eq:leqmax} to be obeyed at all times in order to
avoid cut-off or saturation (for simplicity's aske, the Early geometric representation is here understood
to cover the whole first quadrant, so that saturation and cut-off occurs at the $V_C$ and $I_C$ axes,
respectively.

\begin{eqnarray}
   I \geq 0    \label{eq:geq0} \\
   I \leq  \frac{V_{CC}}{R + R_f}   \label{eq:leqmax}
\end{eqnarray}

As already observed, despite the seeming simplicity of the system of  equations~\ref{eq:system}, the behavior of 
the common emitter configuration turns out to be relatively complex.   However, 
Equations~\ref{eq:I} and~\ref{eq:Ai} can be conveniently used for obtaining almost any information about the
common emitter circuit.  In this work, we will focus on two particularly characteristics of this circuit, namely its 
AC voltage and current gain in terms of the average and coefficient of variation of these gains.  

First, we obtain the expressions for the AC current gain.   This is done by defining the reference $I_B$ and $I$
values for the quiescent point $Q$ as:

\begin{eqnarray}
   I_{B,Q} = I_B(v_i(t) = 0)\\ 
   I_{Q} = I(v_i(t) = 0) 
\end{eqnarray}

so that:

\begin{equation}
   a_i(v_i(t))  = \frac{I(t)  - I_{Q}}{I_B(t) - I_{B,Q}} 
\end{equation}

where $I_B(t)$ and $I(t)$ can be obtained from Equations~\ref{eq:I} and~\ref{eq:Ai}.

The two indices of interest can now be determined by calculating the average and standard
deviation of $a_i$, i.e.~$\langle a_i \rangle$ and $\sigma_{a_i}$.

The voltage gain on the load can now be directly expressed as:

\begin{equation}
   a_v (v_i) = \frac{(I - I_Q) R}{v_i}
\end{equation}

Similarly as developed above, this voltage gain can be characterized in terms of its average
and standard distribution values $\langle a_v \rangle$ and $\sigma_{a_v}$.  The respective
coefficient are particularly relevant, and are henceforth adopted, as they  provide a relative 
quantification of the gain variation, i.e.:

\begin{eqnarray}
   c_{a_i} = \frac{\sigma_{a_i}}{\langle a_i \rangle}   \\
   c_{a_v} = \frac{\sigma_{a_v}}{\langle a_v \rangle}   
\end{eqnarray}

The next sections study the behavior the current and voltage gains for different
circuit (Section~\ref{sec:circ}) and transistor parameter (Section~\ref{sec:trans}) configurations.  It is henceforth 
adopted that  $V_{CC} = 20V$, $R_x = 50000 \Omega$, $R_y = 100000 \Omega$, $R_r = 30 \Omega$, $V_r = 0.6V$, 
and $V = 4V$.

\section{Analysis in Terms of Circuit Parameters}  \label{sec:circ}

In this section we apply the obtained equations describing the operation of the classic common
emitter amplifier while keeping the transistor parameters $V_a$ and $s$ constant and varying
the load resistance $R$ and the feedback resistance $R_f$.   Recall that the gain variation
provides an accurate quantification of the transistor  amplification non-linearity, as larger relative
variations necessarily imply larger distortions and more intense non-linearity.

In order to derive interpretations for typical small signal NPN and PNP silicon BJT transistors,
we use two respective transistor parameters configurations that can often be found experimentally
in real-world silicon BJTs~\cite{costaearly:2018}, i.e.~$(V_{a,npn} = -150 V, s_{npn} =2.5)$ 
and $(V_{a,pnp} = -50, s_{pnp} =10)$.  As the current gain can be approximated as $sV_a$,
these two configurations yield about the same gain, but have widely varying parameters $V_a$
and $s$.

\begin{figure*}[h!]
\centering{
\includegraphics[width=7cm]{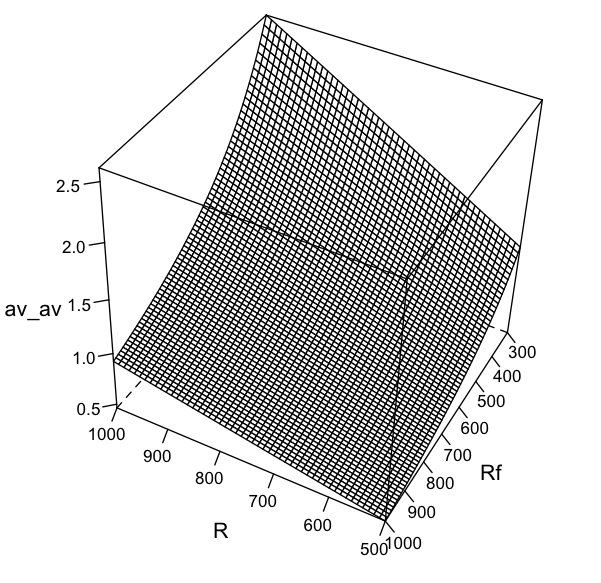}
\includegraphics[width=7cm]{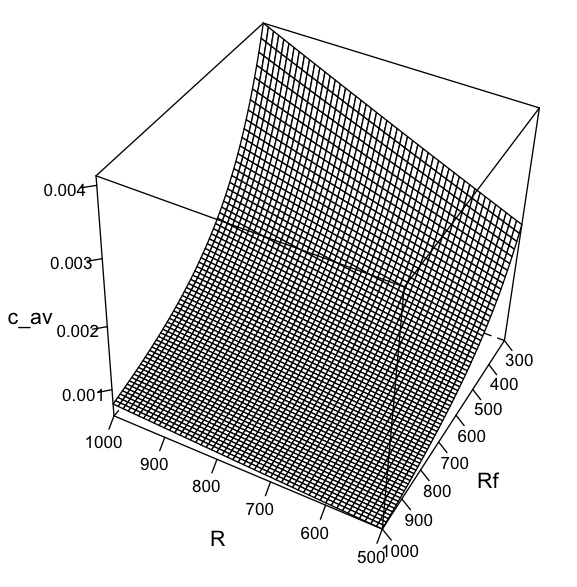}   \\
\includegraphics[width=7cm]{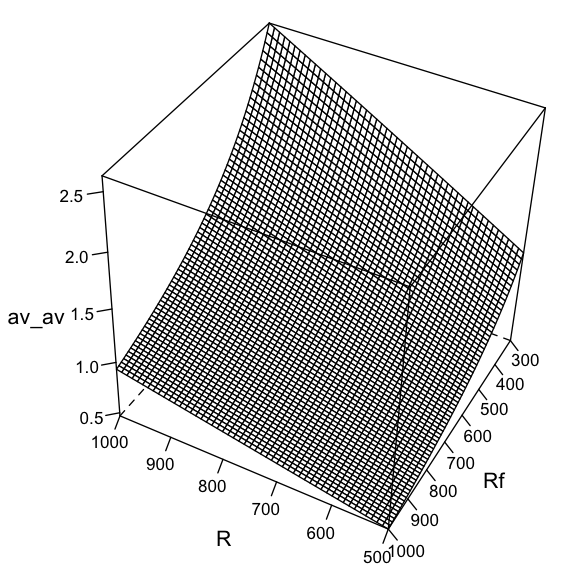}   
\includegraphics[width=7cm]{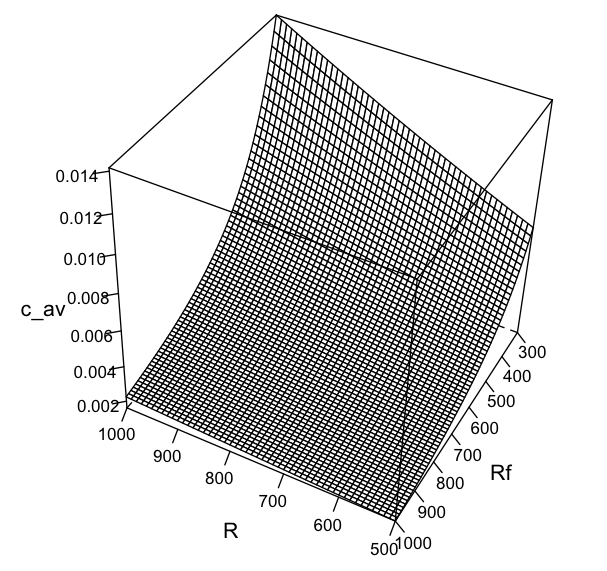}
\caption{The voltage gain average $\langle a_v \rangle$ and coefficient of variation $c_{a_v}$ in terms
of $R$ and $R_f$ obtained
for typical NPN (a-b) and PNP (c-d) configurations of the parameters $V_a$ and $s$.  For NPN, the 
average voltage gain is given in (a), and the respective coefficient of variation in (b), following an analogous
presentation for the PNP case (c-d). }
\label{fig:circ}}
\end{figure*}

Figure~\ref{fig:circ} shows the average AC current gain $\langle a_i \rangle$ (a) as well as the
coefficient of variation of the AC voltage gain $a_v$ (b) with respect to the above mentioned NPN
(a-b) and PNP (c-d) configurations. 

As expected, the two considered parameterizations have nearly the same average gain (Figs.~\ref{fig:circ}(a) and (c)).
Observe that the values given by the respective surfaces tend to agree with the less precise estimation of the 
gain as $R/R_f$  (note that the quiescent current has been chosen as half the $V_{CC}$ value).
The coefficients of variation obtained for the NPN and PNP cases, however, show a marked difference of intensities,
despite the fact that the respective surfaces (i.e.~Fig.~\ref{fig:circ}(b) and (d)) have very similar shapes.  Indeed, the
NPN device has over than 3 times less distortion, as a consequence of higher $V_a$ magnitudes typically found in NPN types.  

The coefficient of variation of the voltage gain correlates strongly (positively) with the average voltage gain. Both
the amplification magnitude and non-linearity increase steadily with $R$ and $R_f$, though tending to increase
faster with the latter.

\section{Analysis in Terms of Transistor Parameters}  \label{sec:trans}

\begin{figure*}[h!]
\centering{
\includegraphics[width=7cm]{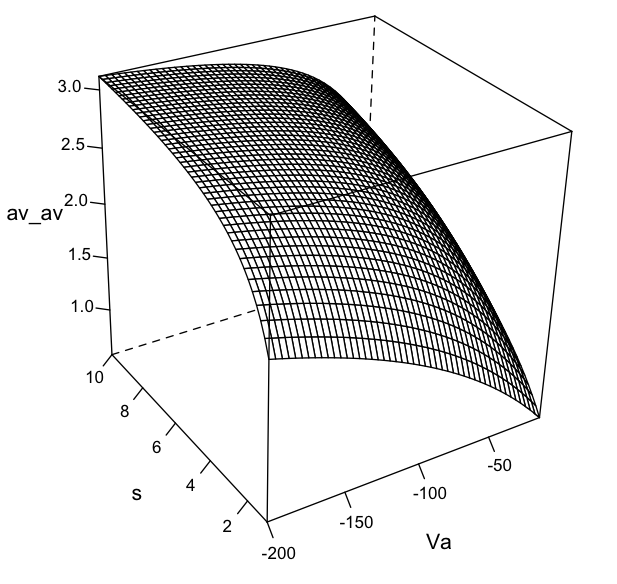}
\includegraphics[width=7cm]{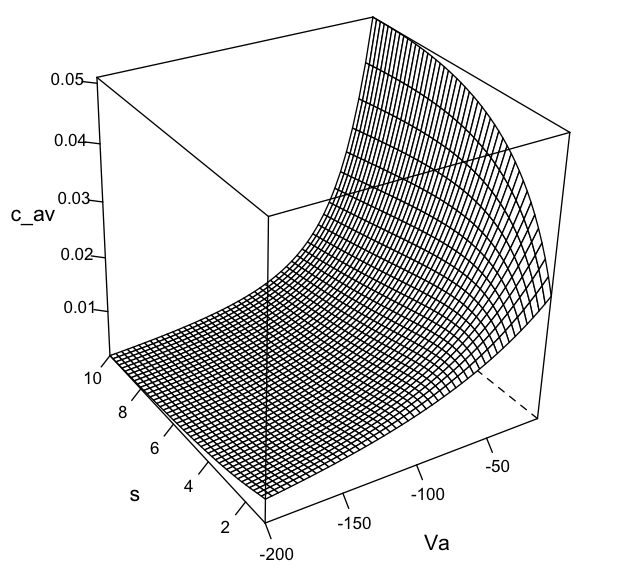}   \\
\includegraphics[width=7cm]{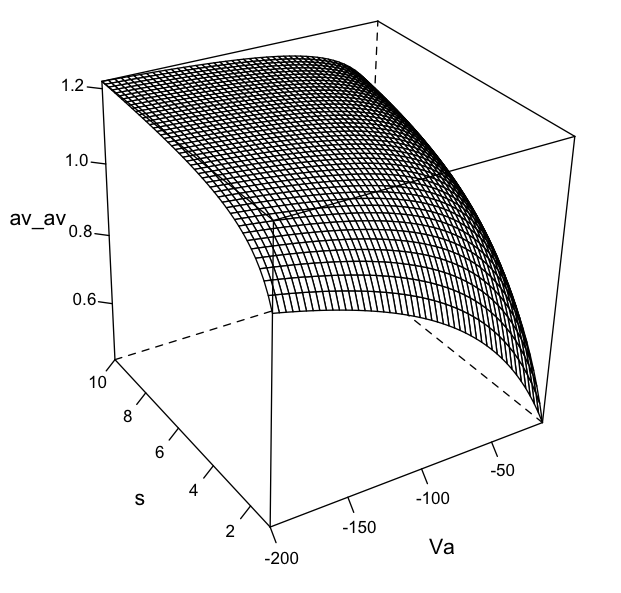}   
\includegraphics[width=7cm]{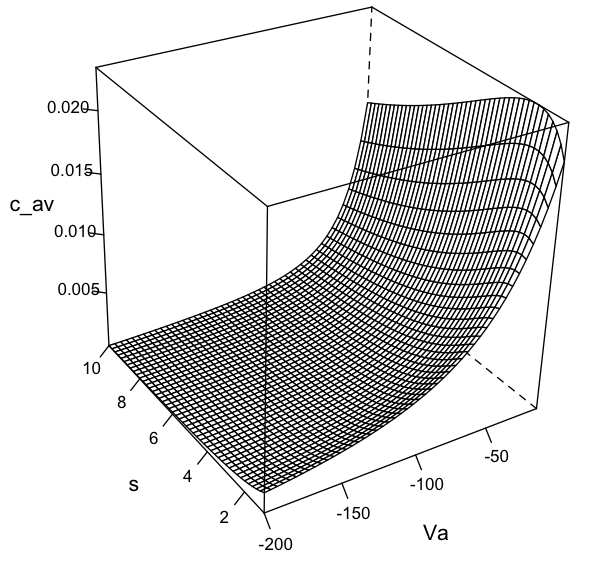}
\caption{The voltage gain average and coefficient of variation in terms of $V_a$ and $s$ for two configurations 
$R = 1000 \Omega$ and $R_f = 300 \Omega$ (a,b) and $R_f = 800 \Omega$ (c,d). See text for discussion.}
\label{fig:circ_trans}}
\end{figure*}

In this section, instead of investigating the behavior of the voltage gain with the circuit parameters $R$ and $R_f$, we
fix $R = 1000\Omega$, take two feedback configurations $R_f = 300 \Omega$ and $800 \Omega$, and systematically vary the Early
model parameters $V_a$ and $s$.   Figure~\ref{fig:circ_trans} presents the respectively obtained results regarding 
the averages and coefficients of variation of the voltage gain throughout the Early parameters space $(-200 \leq V_a \leq -5, 
1 \leq s \leq 10)$.

The average voltage gain surfaces resulted with similar shapes, but showing different magnitudes (Figs.~\ref{fig:circ_trans}, with the 
amplification achieved for the NPN circuit being twice as much higher (approximately $R/R_f$ at the highest values).  
A closer inspection reveals that the average gain surface for $R=1000 \Omega$ and $R_f = 800 \Omega$
has a flatter top than for the other configuration.  These surfaces show that the average voltage gain increases
steeply with both $V_a$ magnitude and $s$, but then reach a plateau of greater gain uniformity  moving toward $(V_a = -200, s= 10)$.

Interestingly, the dispersions of voltage gain (and hence larger non-linearity) verified for 
$R = 1000 \Omega$ and $R_f = 300 \Omega$, as shown in Fig.~\ref{fig:circ_trans}(b), present an opposite 
tendency to that of the average voltage gain.   More specifically, the distortion decreases with both $V_a$ and $s$, 
with this decrease being markedly more accentuated with increase of the $V_a$ magnitude.  A well-defined flat valley of 
linearity is obtained around $(V_a = -200, s= 10)$.  Observe that this valley mostly coincides with the gain peak plateau,
suggesting that the available open loop gain has been traded for increased linearity.

The coefficients of variation obtained for the case $R = 1000 \Omega$ and $R_f = 300 \Omega$ (Fig.~\ref{fig:circ_trans} (d))
resulted distinct  in the sense that the distortion mostly decreases as $s$ becomes larger.  This has the interesting 
implication that transistors with larger values of $s$ will tend, for such situations, to yield less distortion for this specific
configuration of  $R$ and $R_f$.  

Perhaps the most important overall implication of the above discussed results is the fact that, even in presence of relatively
high feedback levels (e.g.~$R_f = 800 \Omega$), when almost all gain is traded-off for linearity and stability, the linearity of the
amplification still varied \emph{substantially} with different values of the transistor parameters $V_a$ and $s$.  
This finding corroborates, for the considered configuration and ranges, the previously observed 
phenomenon~\cite{costafeed:2017,costaearly:2017}
that negative feedback may not be capable of completely eliminating transistor parameter variation.   Therefore, it
becomes an interesting option to consider the transistor parameters $V_a$ and $s$ for circuit design and implementation,
even when adopting intense levels of negative feedback.

\section{Current Gain}

Though the common emitter amplifier is typically approached with respect to voltage gain, It is also interesting to consider
the AC current gain as can be derived from the analytical Equations~\ref{eq:I} and~\ref{eq:Ai}.  

\begin{figure}[h!]
\centering{
\includegraphics[width=7cm]{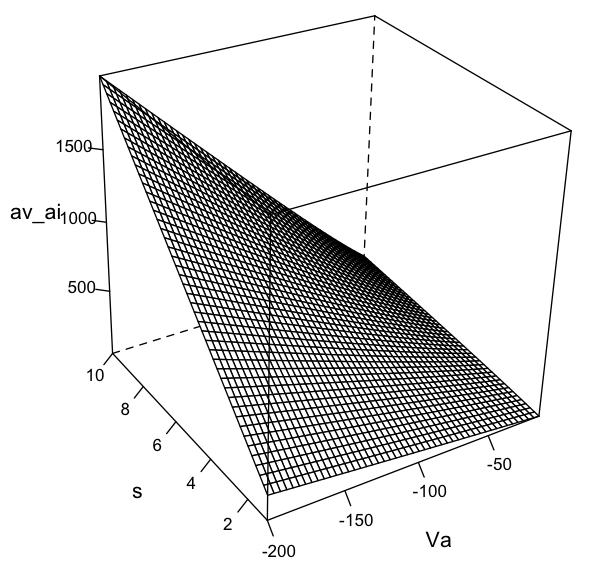}
\caption{ }
\label{fig:ai}}
\end{figure}

The obtained results are qualitatively similar (though with distinct intensities), so we restrict our discussion to the analysis of 
the average current gain throughout the space $(V_a,s)$, as observed for $R = 1000 \Omega$ and $R_f = 300 \Omega$, 
which is depicted in Figure~\ref{fig:ai}.

This result shows that the average current gain for the mentioned configuration increases in an almost linear way
with both $V_a$ and $s$, reaching its peak of nearly 2000 at $(V_a = -200, s= 10)$. Though a direct consequence
of the circuit operation dynamics, it is interesting to observe that the average current and voltage gains depend in
such distinct ways on the parameters $V_a$ and $s$.

\section{Concluding Remarks}

Amplification is a key issue in analog/linear electronics because of the non-linearity that is inherent to any real-world
amplifying devices such as transistors.  As  such, much effort has been invested in devising circuits and techniques
capable of achieving satisfactory performance under specific circumstances.  The common emitter circuit is probably the
``quintessential'' amplification approach in electronics, acting as a reference for other designs and development of novel
techniques aimed at improving amplification, as is the case of negative feedback.  

The non-linearity inherent to junction transistors stems mostly from the Early effect, which implies the base current-indexed 
isolines characterizing the typical amplification to converge into a single point along the $V_C$ axis.   The recent introduction 
of the Early modeling  approach~\cite{costaearly:2017,costaearly:2018,germanium:2018,costaequiv:2018} paved the way to 
more systematic, analytical investigations of the behavior of amplifying circuits, including the common emitter configuration.  
In addition to its inherent simplicity, these approaches are capable of taking in to account the transistor non-linearity as
stemming from the radiating characteristic isolines implied by the Early effect.
Preliminary approaches to common emitter amplification were reported recently~\cite{costaequiv:2018,costareact:2018}, 
but those approaches took into account a simplified common emitter configuration devoid of negative feedback.  Even so, 
it was possible to achieve several interesting results, including the relatively severe effect of the transistor non-linear 
amplification, as well as its great dependence on the transistor Early parameters $V_a$ and $s$.  

The present work addressed the complete, classic common emitter configuration, incorporating voltage divider-based
input biasing and, more importantly, negative feedback as provided by a resistance $R_f$ attached between the collector
and the ground.  This resistance has to be kept at relatively small values in order not to reduce too much the current and
voltage amplification too much.  The approach reported here takes into account the transistor non-linearity implied by
the Early effect, allowing the study of how effectively negative feedback can enhance linearity.
 
The reported analysis was developed as follows.   First, the equivalent circuit of the classic common emitter configuration
was derived by using electrical circuit laws and the Early voltage equivalent circuit for a transistor in amplifying mode.
The base current $I_B$ and collector current $I$ were treated as functions of the input voltage $v_i(t)$.  Analytical
solutions were obtained for $I_B$ and $I$, providing the complete electrical description of the considered common 
emitter amplifier.  The average, standard deviation, and coefficient of variation of the voltage and, to a smaller extent,
current amplification (gains) were obtained and used for investigating the effect of several circuit and transistor parameter 
configurations. As gain variations are closely related to non-linearities, it was possible to address both the overall gain, as expressed
in terms of the average of the gains obtained along $v_i(t)$ excursions, as well as to infer the respective linearities.

Several interesting results have been reported.   First, we have that the choices of the load and feedback resistance (circuit
components) can influence strongly both average gain (as expected) and linearity.   Importantly, these 
gains were found to vary with $v_i(t)$ even in presence of relatively intense negative feedback, implying amplifying distortion.  
Perhaps more importantly, the transistor inherent characteristics $V_a$ and $s$ were found to strongly influence
the linearity of the amplification \emph{even in presence of relatively high negative feedback}.  Similar results had been
experimentally observed recently~\cite{costafeed:2017,costaearly:2018}.  By providing a more complete and accurate
analytical framework to characterize the common emitter circuit, the present work not only confirmed those preliminary 
experimental  indications, but also showed in a more systematic and conclusive way that, though certainly remaining an 
interesting option, \emph{negative feedback is not enough to completely eliminate effects of transistor parameter variations}.  
This important fact implies that, especially in particularly critical circumstances, \emph{it may be worth taking into account the specific 
characteristics of available real-world transistors.}  Another interesting result concerns the fact that the linearity can increase
or decrease according to $s$ for different $R$ and $R_f$ configurations.  
 
The implications of the above results are several and have great potential for motivating many related further
investigations, including the study of higher power amplifiers, consideration of reactive loads, as well as the systematic
study and characterization of other important electronic circuits such as current mirror, phase splitters, and differential 
amplifiers, up to complete operational amplifier configurations, to name but a few possibilities.

\vspace{0.7cm}
\textbf{Acknowledgments.}

Luciano da F. Costa
thanks CNPq (grant no.~307333/2013-2) for sponsorship. This work has benefited from
FAPESP grants 11/50761-2 and 2015/22308-2.
\vspace{1cm}


\bibliography{mybib}
\bibliographystyle{plain}
\end{document}